\documentclass{article}
\usepackage{spconf,amsmath,graphicx}
\usepackage{color}

\usepackage{wasysym}
\usepackage{dblfloatfix}
\usepackage{multirow,tabularx}
\usepackage{booktabs}
\usepackage{caption} 
\usepackage{setspace}
\title{Crepe: A Convolutional Representation for Pitch Estimation}
\name{Jong Wook Kim$^1$, Justin Salamon$^{1,2}$, Peter Li$^1$, Juan Pablo Bello$^1$}
\address{$^1$Music and Audio Research Laboratory, New York University\\
	$^2$Center for Urban Science and Progress, New York University}
\begin{document}
\ninept
\maketitle

\begin{abstract}
	The task of estimating the fundamental frequency of a monophonic sound recording, also known as pitch tracking, is fundamental to audio processing with multiple applications in speech processing and music information retrieval.
	To date, the best performing techniques, such as the pYIN algorithm, are based on a combination of DSP pipelines and heuristics.
	While such techniques perform very well on average, there remain many cases in which they fail to correctly estimate the pitch.
	In this paper, we propose a data-driven pitch tracking algorithm, CREPE, which is based on a deep convolutional neural network that operates directly on the time-domain waveform.
	We show that the proposed model produces state-of-the-art results, performing equally or better than pYIN.
	Furthermore, we evaluate the model's generalizability in terms of noise robustness.
	A pre-trained version of CREPE is made freely available as an open-source Python module for easy application.
\end{abstract}

\keywords{pitch estimation, convolutional neural network}

\section{Introduction}\label{sec:introduction}

Estimating the fundamental frequency (f0) of a monophonic audio signal, also known as pitch tracking or pitch estimation, is a long-standing topic of research in audio signal processing.
Pitch estimation plays an important role in music signal processing, where monophonic pitch tracking is used as a method to generate pitch annotations for multi-track datasets \cite{bittner2014medleydb} or as a core component of melody extraction systems \cite{bosch2014melody, mauch2015computer}. 
Pitch estimation is also important for speech analysis, where prosodic aspects such as intonations may reflect various features of speech \cite{zubizarreta1998prosody}.

Pitch is defined as a subjective quality of perceived sounds and does not precisely correspond to the physical property of the fundamental frequency \cite{hartmann1997signals}.
However, apart from a few rare exceptions, pitch can be quantified using fundamental frequency, and thus they are often used interchangeably outside psychoacoustical studies. 
For convenience, we will also use the two terms interchangeably throughout this paper.

Computational methods for monotonic pitch estimation have been studied for more than a half-century \cite{noll1967cepstrum}, and many reliable methods have been proposed since.
Earlier methods commonly employ a certain candidate-generating function, accompanied by pre- and post-processing stages to produce the pitch curve.
Those functions include the cepstrum \cite{noll1967cepstrum}, the autocorrelation function (ACF) \cite{dubnowski1976acf}, the average magnitude difference function (AMDF) \cite{ross1974amdf}, the normalized cross-correlation function (NCCF) as proposed by RAPT \cite{talkin1995rapt} and PRAAT \cite{boersma1993praat}, and the cumulative mean normalized difference function as proposed by YIN \cite{de2002yin}. More recent approaches include SWIPE \cite{camacho2008swipe}, which performs template matching with the spectrum of a sawtooth waveform, and 
pYIN \cite{mauch2014pyin}, a probabilistic variant of YIN that uses a Hidden Markov Model (HMM) to decode the most probable sequence of pitch values.
According to a few comparative studies, the state of the art is achieved by YIN-based methods \cite{von2010comparison, babacan2013comparative}, with pYIN being the best performing method to date \cite{mauch2014pyin}.

A notable trend in the above methods is that the derivation of a better pitch detection system solely depends on cleverly devising a robust candidate-generating function and/or sophisticated post-processing steps, i.e.~heuristics, and none of them are directly learned from data, except for manual hyperparameter tuning.
This contrasts with many other problems in music information retrieval like chord ID \cite{humphrey2012rethinking} and beat detection \cite{bock2011enhanced}, where data-driven methods have been shown to consistently outperform heuristic approaches.
One possible explanation for this is that since fundamental frequency is a low-level physical attribute of an audio signal which is directly related to its periodicity, in many cases heuristics for estimating this periodicity perform extremely well with accuracies (measured in raw pitch accuracy, defined later on) close to 100\%, leading some to consider the task a solved problem.
This, however, is not always the case, and even top performing algorithms like pYIN can still produce noisy results for challenging audio recordings such as a sound of uncommon instruments or a pitch curve that fluctuates very fast.
This is particularly problematic for tasks that require a flawless f0 estimation, such as using the output of a pitch tracker to generate reference annotations for melody and multi-f0 estimation \cite{salamon2017analysis,bittner2017deepsalience}.

In this paper, we present a novel, data-driven method for monophonic pitch tracking based on a deep convolutional neural network operating on the time-domain signal.
We show that our approach, CREPE (Convolutional  Representation  for  Pitch  Estimation), obtains state-of-the-art results, outperforming heuristic approaches such as pYIN and SWIPE while being more robust to noise too.
We further show that CREPE is highly precise, maintaining over 90\% raw pitch accuracy even for a strict evaluation threshold of just 10 cents.
The Python implementation of our proposed approach, along with a pre-trained model of CREPE are made available online\footnote{\texttt{https://github.com/marl/crepe}} for easy utilization and reproducibility.

\section{Architecture}

\setlength{\belowcaptionskip}{-5pt}
\captionsetup[figure]{skip=5pt}
\begin{figure*}
	\includegraphics[width=\textwidth]{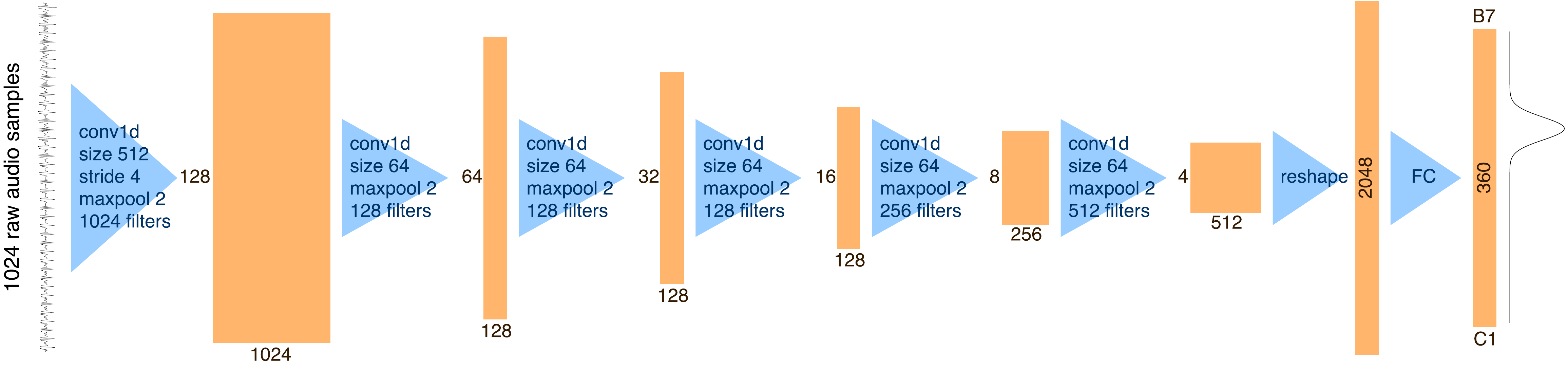}
	\caption{The architecture of the CREPE pitch tracker. The six convolutional layers operate directly on the time-domain audio signal, producing an output vector that approximates a Gaussian curve as in Equation \ref{eqn:gaussian}, which is then used to derive the exact pitch estimate as in Equation \ref{eqn:resulting}.}
	\label{fig:architecture}
\end{figure*}

CREPE consists of a deep convolutional neural network which operates directly on the time-domain audio signal to produce a pitch estimate.
A block diagram of the proposed architecture is provided in Figure \ref{fig:architecture}.
The input is a 1024-sample excerpt from the time-domain audio signal, using a 16 kHz sampling rate.
There are six convolutional layers that result in a 2048-dimensional latent representation, which is then connected densely to the output layer with sigmoid activations corresponding to a 360-dimensional output vector $\hat{\mathbf{y}}$.
From this, the resulting pitch estimate is calculated deterministically.

Each of the 360 nodes in the output layer corresponds to a specific pitch value, defined in cents.
Cent is a unit representing musical intervals relative to a reference pitch $f_{\mathrm{ref}}$ in Hz, defined as a function of frequency $f$ in Hz:
\begin{equation}
	\cent(f) = 1200 \cdot \log_2 \frac{f}{f_{\mathrm{ref}}},
\end{equation}
where we use $f_{\mathrm{ref}} = 10 \mathrm{~Hz}$ throughout our experiments. 
This unit provides a logarithmic pitch scale where 100 cents equal one semitone.
The 360 pitch values are denoted as $\cent_1, \cent_2, \cdots, \cent_{360}$ and are selected so that they cover six octaves with 20-cent intervals between C1 and B7, corresponding to 32.70 Hz and 1975.5 Hz. 
The resulting pitch estimate $\hat{\cent}$ is the weighted average of the associated pitches $\cent_i$ according to the output $\hat{\mathbf{y}}$, which gives the frequency estimate in Hz:
\begin{equation}\label{eqn:resulting}
\hat{\cent} = \frac{\sum_{i=1}^{360}\hat{y}_i \cent_i}{\sum_{i=1}^{360} \hat{y}_i}, ~~~~~~~~~~~~~~
\hat{f} = f_{\mathrm{ref}} \cdot 2 ^ {\hat{\cent} / 1200}.
\end{equation}

The target outputs we use to train the model are 360-dimensional vectors, where each dimension represents a frequency bin covering 20 cents (the same as the model's output).
The bin corresponding to the ground truth fundamental frequency is given a magnitude of one.
As in \cite{bittner2017deepsalience}, in order to soften the penalty for near-correct predictions, the target is Gaussian-blurred in frequency such that the energy surrounding a ground truth frequency decays with a standard deviation of 25 cents:
\begin{equation}\label{eqn:gaussian}
	y_i = \exp \left ( {-\frac{(\cent_i - \cent_{\mathrm{true}})^2}{2 \cdot 25^2}} \right ),
\end{equation}
This way, high activations in the last layer indicate that the input signal is likely to have a pitch that is close to the associated pitches of the nodes with high activations.

The network is trained to minimize the binary cross entropy between the target vector $\mathbf{y}$ and the predicted vector $\mathbf{\hat{y}}$:
\begin{equation}
\mathcal{L}(\mathbf{y}, \mathbf{\hat{y}}) = \sum_{i=1}^{360} \left ( - y_i \log \hat{y_i} - (1 - y_i) \log (1 - \hat{y_i}) \right ),
\end{equation}
where both $y_i$ and $\hat{y}_i$ are real numbers between 0 and 1.
This loss function is optimized using the ADAM optimizer \cite{kingma2015adam}, with the learning rate 0.0002. 
The best performing model is selected after training until the validation accuracy no longer improves for 32 epochs, where one epoch consists of 500 batches of 32 examples randomly selected from the training set. 
Each convolutional layer is preceded with batch normalization \cite{ioffe2015batch} and followed by a dropout layer \cite{srivastava2014dropout} with the dropout probability 0.25.
This architecture and the training procedures are implemented using Keras \cite{chollet2015keras}.

\section{Experiments}

\subsection{Datasets}

In order to objectively evaluate CREPE and compare its performance to alternative algorithms, we require audio data with perfect ground truth annotations.
This is especially important since the performance of the compared algorithms is already very high.
In light of this, we cannot use a dataset such as MedleyDB \cite{bittner2014medleydb}, since its annotation process includes manual corrections which do not guarantee a 100\% perfect match between the annotation and the audio, and it can be affected, to a degree, by human subjectivity.
To guarantee a perfectly objective evaluation, we must use datasets of synthesized audio in which we have perfect control over the f0 of the resulting signal.
We use two such datasets: the first, RWC-synth, contains 6.16 hours of audio synthesized from the RWC Music Database \cite{goto2002rwc} and is used to evaluate pYIN in \cite{mauch2014pyin}.
It is important to note that the signals in this dataset were synthesized using a fixed sum of a small number of sinusoids, meaning that the dataset is highly homogenous in timbre and represents an over-simplified scenario.
To evaluate the algorithms under more realistic (but still controlled) conditions, the second dataset we use is a collection of 230 monophonic stems taken from MedleyDB and re-synthesized using the methodology presented in \cite{salamon2017analysis}, which uses an analysis/synthesis approach to generate a synthesized track with a perfect f0 annotation that maintains the timbre and dynamics of the original track.
This dataset consists of 230 tracks with 25 instruments, totaling 15.56 hours of audio, and henceforth referred to as MDB-stem-synth.

%\captionsetup[figure]{skip=0pt}
\begin{figure*}[b!]
	\includegraphics[width=\textwidth]{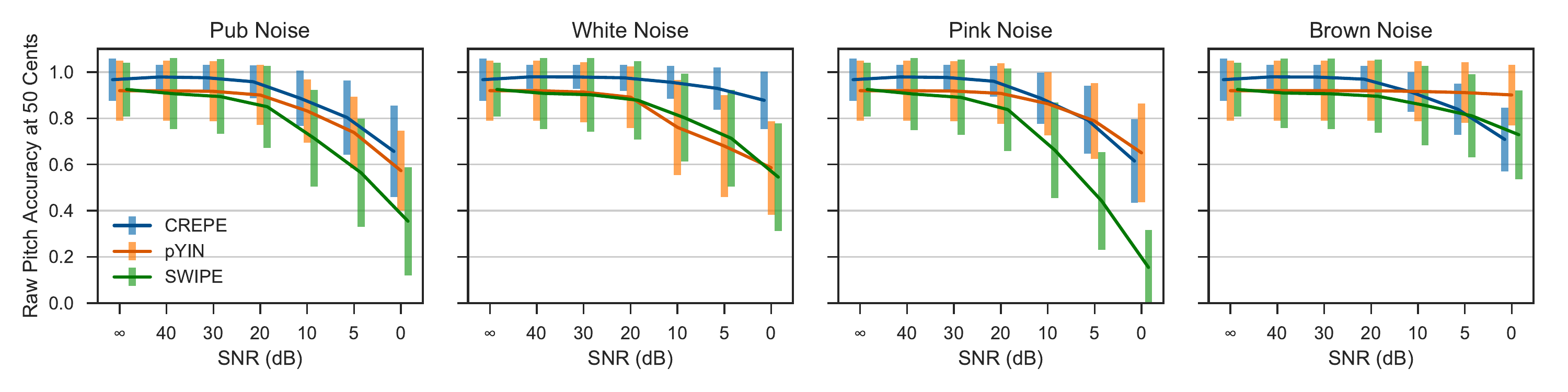}
	\caption{Pitch tracking performance when additive noise signals are present. The error bars are centered at the average raw pitch accuracies and span the first standard deviations. With brown noise being a notable exception, CREPE shows the highest noise robustness in general. }
	\label{fig:noise}
\end{figure*}

\subsection{Methodology}

We train the model using 5-fold cross-validation, using a 60/20/20 train, validation, and test split.
For MDB-stem-synth, we use artist-conditional folds, in order to avoid training and testing on the same artist which can result in artificially high performance due to artist or album effects \cite{sturm2013classification}.
The evaluation of an algorithm's pitch estimation is measured in raw pitch accuracy (RPA) and raw chroma accuracy (RCA) with 50 cent thresholds \cite{salamon2014melody}.
These metrics measure the proportion of frames in the output for which the output of the algorithm is within 50 cents (a quarter-tone) of the ground truth.
We use the reference implementation provided in \texttt{mir\_eval} \cite{raffel2014mir_eval} to compute the evaluation metrics.

We compare CREPE against the current state of the art in monophonic pitch tracking, represented by the pYIN \cite{mauch2014pyin} and SWIPE \cite{camacho2008swipe} algorithms.
To examine the noise robustness of each algorithm, we also evaluate their pitch tracking performance on degraded versions of MDB-stem-synth, using the Audio Degradation Toolbox (ADT) \cite{mauch2013adt}.
We use four different noise sources provided by the ADT: pub, white, pink, and brown.
The pub noise is an actual recording of the sound in a crowded pub, 
and the white noise is a random signal with a constant power spectral density over all frequencies.
The pink and brown noise have the highest power spectral density in low frequencies, and the densities fall off at 10 dB and 20 dB per decade respectively.
We used seven different signal-to-noise ratio (SNR) values: $\infty$, 40, 30, 20, 10, 5, and 0 dB.

\subsection{Results}

\subsubsection{Pitch Accuracy}

Table \ref{tbl:accuracy50} shows the pitch estimation performance tested on the two datasets.
On the RWC-synth dataset, CREPE yields a close-to-perfect performance where the error rate is lower than the baselines by more than an order of magnitude.
While these high accuracy numbers are encouraging, those are achievable thanks to the highly homogeneous timbre of the dataset.
In order to test the generalizability of the algorithms on a more timbrally diverse dataset, we evaluated the performance on the MDB-stem-synth dataset as well.
It is notable that the degradation of performance from RWC-synth is more significant for the baseline algorithms, implying that CREPE is more robust to complex timbres compared to pYIN and SWIPE.

Finally, to see how the algorithms compare under scenarios where any deviation in the estimated pitch from the true value could be detrimental, in Table \ref{tbl:thresholds} we report the RPA at lower evaluation tolerance thresholds of 10 and 25 cents as well as the RPA at the standard 50 cents threshold for reference.
We see that as the threshold is decreased, the difference in performance becomes more accentuated, with CREPE 
%maintaining near-perfect RPA for RWC-synth while the others degrade significantly.
 outperforming by over 8 percentage points when the evaluation tolerance is lowered to 10 cents.
This suggests that CREPE is especially preferable when even minor deviations from the true pitch should be avoided as best as possible.
Obtaining highly precise pitch annotations is perceptually meaningful for transcription and analysis/resynthesis applications.

\subsubsection{Noise Robustness}

Noise robustness is key to many applications like speech analysis for mobile phones or smart speakers, or for live music performance.
In Figure \ref{fig:noise} we show how the pitch estimation performance is affected when an additive noise is present in the input signal.
CREPE maintains the highest accuracy for all SNR levels for pub noise and white noise, and for all SNR levels except for the highest level of pink noise.
Brown noise is the exception where pYIN's performance is almost unaffected by the noise.
This can be attributed to the fact that brown noise has most of its energy at low frequencies, to which the YIN algorithm (on which pYIN is based) is particularly robust.

To summarize, we confirmed that CREPE performs better in all cases where the SNR is below 10 dB while the performance varies depending on the spectral properties of the noise when the noise level is higher, which indicates that our approach can be reliable under a reasonable amount of additive noise.
CREPE is also more stable, exhibiting consistently lower variance in performance compared to the baseline algorithms.

\setlength{\belowcaptionskip}{-0pt}
\begin{table}[t]
	\begin{center}
		\setlength{\tabcolsep}{0.5em}
		\renewcommand{\arraystretch}{1.3}
		\begin{tabular}{c|c||c|c|c} \hline
			\multicolumn{1}{c}{Dataset} & \multicolumn{1}{c}{Metric} & \multicolumn{1}{c}{CREPE} & \multicolumn{1}{c}{pYIN} & \multicolumn{1}{c}{SWIPE} \\ \hline
			\renewcommand{\arraystretch}{1.8}
			\multirow{2}{*}{\parbox[c]{1cm}{\centering RWC-synth}} & RPA & \textbf{0.999$\pm$0.002} & 0.990$\pm$0.006& 0.963$\pm$0.023 \\ \cline{2-5}
			& RCA & \textbf{0.999$\pm$0.002} & 0.990$\pm$0.006& 0.966$\pm$0.020 \\ \hline \hline
			\multirow{2}{*}{\parbox[c]{1cm}{\centering \begin{spacing}{0.75} MDB-stem-synth \end{spacing}}} & RPA & \textbf{0.967$\pm$0.091} & 0.919$\pm$0.129& 0.925$\pm$0.116 \\ \cline{2-5}
			& RCA & \textbf{0.970$\pm$0.084} & 0.936$\pm$0.092& 0.936$\pm$0.100 \\ \hline
		\end{tabular}
	\end{center}
	\caption{Average raw pitch/chroma accuracies and their standard deviations, tested with the 50 cents threshold.}
	\label{tbl:accuracy50}
\end{table}

\begin{table}[t]
	\begin{center}
		\setlength{\tabcolsep}{0.4em}
		\renewcommand{\arraystretch}{1.3}
		\begin{tabular}{c|c||c|c|c} \hline
			\multicolumn{1}{c}{Dataset} & \multicolumn{1}{c}{Threshold} &  \multicolumn{1}{c}{CREPE} & \multicolumn{1}{c}{pYIN} & \multicolumn{1}{c}{SWIPE} \\ \hline
			
			\multirow{3}{*}{\parbox[c]{1cm}{\centering RWC-synth}} 
			& 50 cents & \textbf{0.999$\pm$0.002} & 0.990$\pm$0.006 & 0.963$\pm$0.023 \\ \cline{2-5}
			& 25 cents & \textbf{0.999$\pm$0.003} & 0.972$\pm$0.012 & 0.949$\pm$0.026 \\ \cline{2-5}
			& 10 cents & \textbf{0.995$\pm$0.004} & 0.908$\pm$0.032 & 0.833$\pm$0.055 \\ \hline \hline
			
			\multirow{3}{*}{\parbox[c]{1cm}{\centering MDB-stem-synth}} 
			& 50 cents & \textbf{0.967$\pm$0.091} & 0.919$\pm$0.129 & 0.925$\pm$0.116 \\ \cline{2-5}
			& 25 cents & \textbf{0.953$\pm$0.103} & 0.890$\pm$0.134 & 0.897$\pm$0.127 \\ \cline{2-5}
			& 10 cents & \textbf{0.909$\pm$0.126} & 0.826$\pm$0.150 & 0.816$\pm$0.165 \\ \hline
		\end{tabular}
	\end{center}
	\caption{Average raw pitch accuracies and their standard deviations, with different evaluation thresholds.}
	\label{tbl:thresholds}
\end{table}

\setlength{\belowcaptionskip}{-8pt}
\captionsetup[figure]{skip=3pt}
\begin{figure*}[t]
	\begin{minipage}{0.48\textwidth}
		\begin{center}
			\includegraphics[width=\columnwidth]{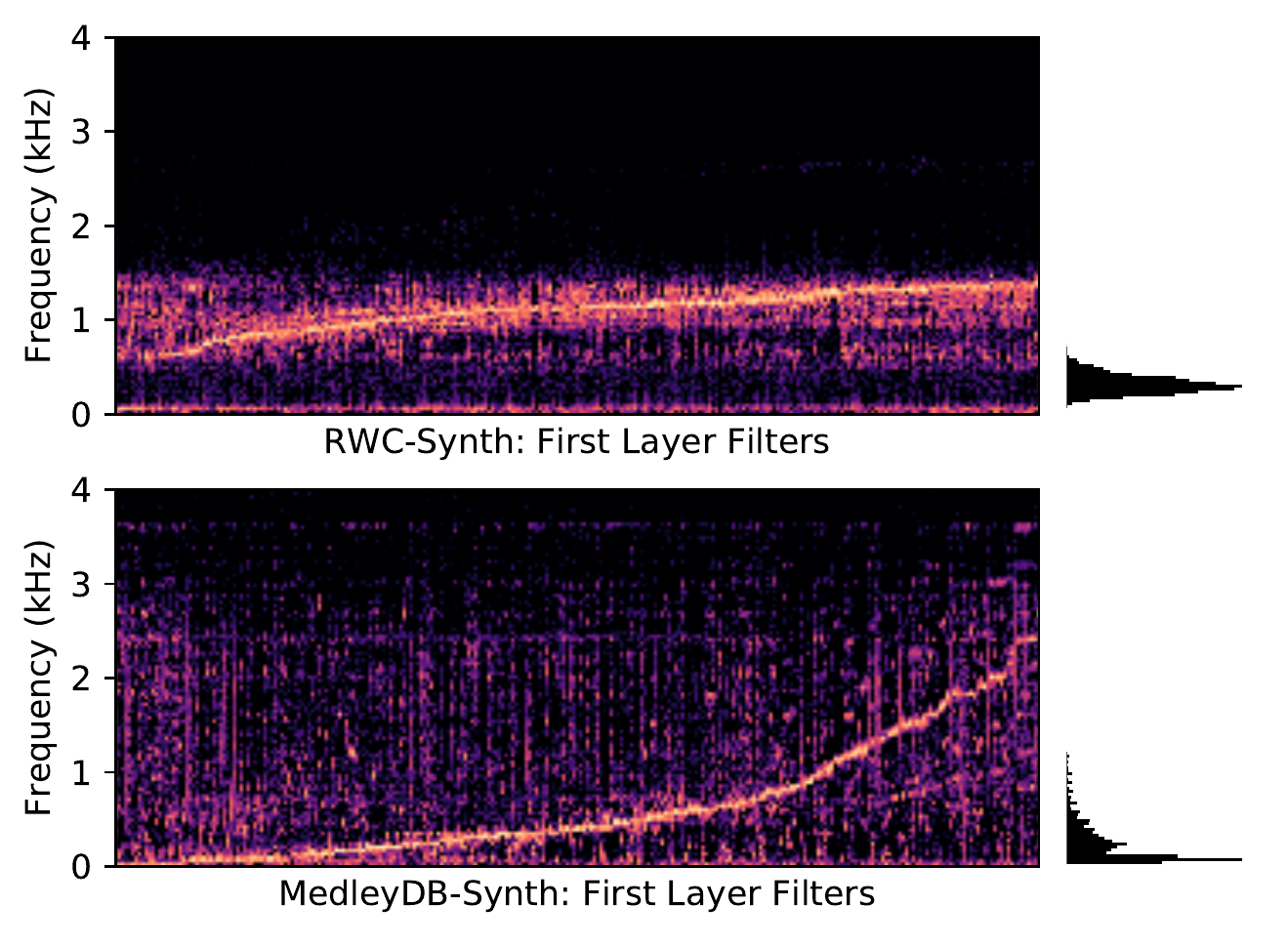}
		\end{center}
		\vspace{-10pt}
		\caption{
			Fourier spectra of the first-layer filters sorted by the frequency of the peak magnitude.
			Histograms on the right show the distribution of ground-truth frequencies in the corresponding dataset.
		}
		\label{fig:firstlayer}
	\end{minipage}
	~~~~~~
	\begin{minipage}{0.48\textwidth}
		\begin{center}
			\includegraphics[width=\columnwidth]{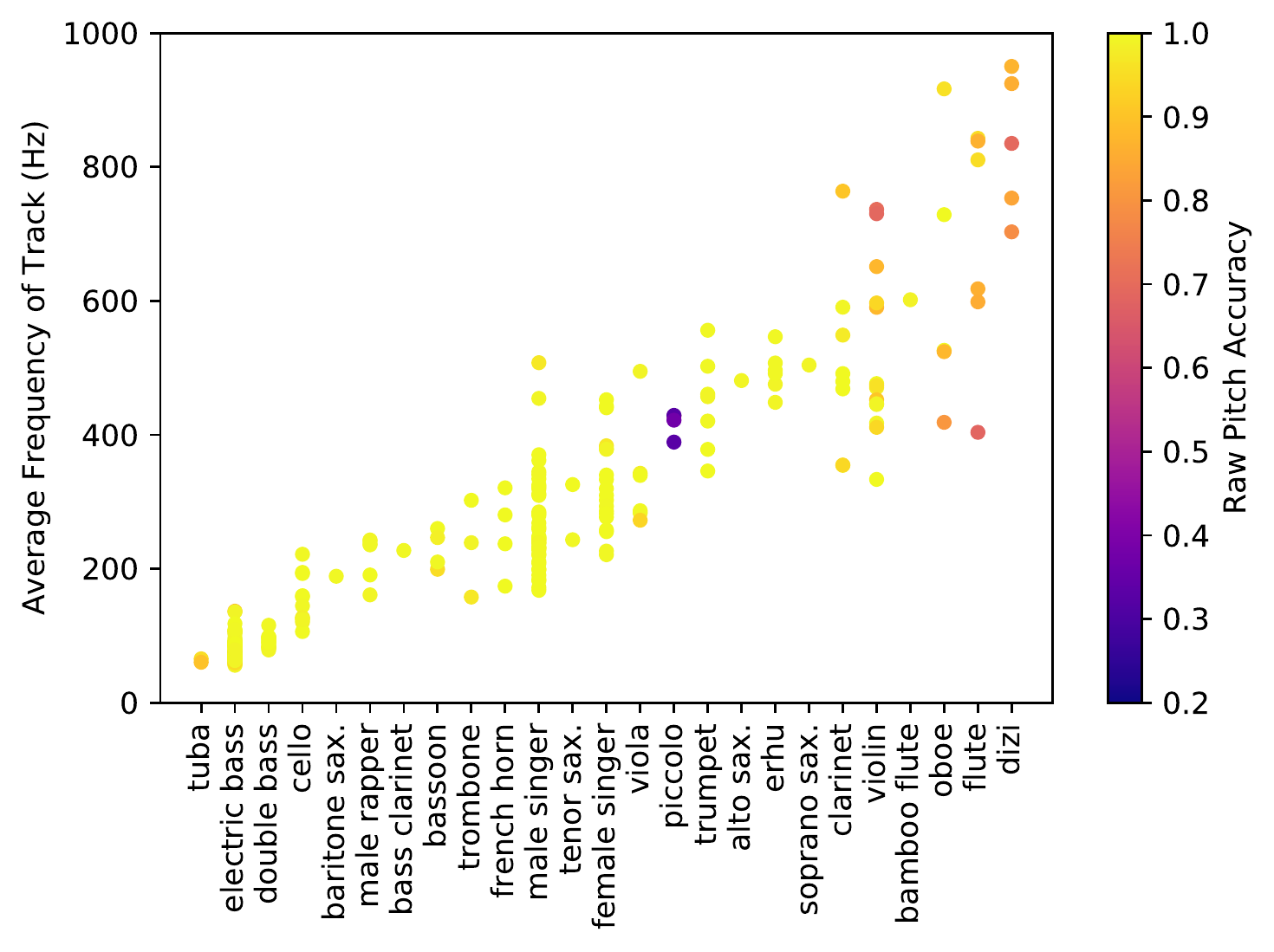}
		\end{center}
		\vspace{-10pt}
		\caption{
			The raw pitch accuracy (RPA) of CREPE's predictions on each of the 230 tracks in MDB-stem-synth with respect to the instrument, sorted by the average frequency.
			% The  low-performing  cases  are  concentrated  in  the extreme frequencies, showing that the model performs better for the sfrequency range that are well-represented in the training set.
		}
		\label{fig:per-track}
	\end{minipage}
\end{figure*}

\subsubsection{Model Analysis}

To gain some insight into the CREPE model, in Figure \ref{fig:firstlayer} we visualize the spectra of the 1024 convolutional filters in the first layer of the neural network, with histograms of the ground-truth frequencies to the right of each plot.
It is noticeable that the filters learned from the RWC-synth dataset have the spectral density concentrated between 600 Hz and 1500 Hz, while the ground-truth frequencies are mostly between 100 Hz and 600 Hz.
This indicates that the first convolutional layer in our model learns to distinguish the frequencies of the overtones rather than the fundamental frequency.
These filters focusing on overtones are also visible for MDB-stem-synth, where peak frequencies of the filters range well above the f0 distribution of the dataset, but in this case, the majority of the filters overlap with the ground-truth distribution, unlike RWC-synth.
A possible explanation for this is that since the timbre in RWC-synth is fixed and identical for all tracks, the model is able to obtain a highly accurate estimate of the f0 by modeling its harmonics.
Conversely, when the timbre is heterogeneous and more complex, as is the case for MDB-stem-synth, the model cannot rely solely on the harmonic structure and requires filters that capture the f0 periodicity directly in addition to the harmonics.
In both cases, this suggests that the neural network can adapt to the distribution of timbre and frequency in the dataset of interest, which in turn contributes to the higher performance of CREPE compared to the baseline algorithms.

\subsubsection{Performance by Instrument}

The MDB-stem-synth dataset contains 230 tracks from 25 different instruments, where electric bass (58 tracks) and male singer (41 tracks) are the most common while there are instruments that occur in only one or two tracks.
In Figure \ref{fig:per-track} we plot the performance of CREPE on each of the 230 tracks, with respect to the instrument of each track.
It is notable that the model performs worse for the instruments with higher average frequencies, but the performance is also dependent on the timbre.
CREPE performs particularly worse on the tracks with the dizi, a Chinese transverse flute, because the tracks came from the same artist, and they are all placed in the same split.
This means that for the fold in which the dizi tracks are in the test set, the training and validation sets do not contain a single dizi track, and the model fails to generalize to this previously unseen timbre.
There are 5 instruments (bass clarinet, bamboo flute, and the family of saxophones) that occur only once in the dataset, but their performance is decent, because their timbres do not deviate too far from other instruments in the dataset.
For the flute and the violin, although there are many tracks with the same instrument in the training set, the performance is low when the sound in the tested tracks is too low (flute) or too high (violin) compared to other tracks of the same instruments.
The low performance on the piccolo tracks is due to an error in the dataset where the annotation is inconsistent with the correct pitch range of the instrument.
Unsurprisingly, the model performs well on test tracks whose timbre and frequency range are well-represented in the training set.

\section{Discussions and Conclusion}

In this paper, we presented a novel data-driven method for monophonic pitch tracking based on a deep convolutional neural network operating on time-domain input, CREPE.
We showed that CREPE obtains state-of-the-art results, outperforming pYIN and SWIPE on two datasets with homogeneous and heterogeneous timbre respectively.
Furthermore, we showed that CREPE remains highly accurate even at a very strict evaluation threshold of just 10 cents.
We also showed that in most cases CREPE is more robust to added noise.

Ideally, we want the model to be invariant to all transformations that do not affect pitch, such as changes due to distortion and reverberation.
Some invariance can be induced by the architectural design of the model, such as the translation invariance induced by pooling layers in our model as well as in deep image classification models.
However, it is not as straightforward to design the model architecture to specifically ignore other pitch-preserving transformations.
While it is still an intriguing problem to build an architecture to achieve this, we could use data augmentation to generate transformed and degraded inputs that can effectively make the model learn the invariance.
The robustness of the model could also be improved by applying pitch-shifts as data augmentation \cite{mcfee2015muda} to cover a wider pitch range for every instrument.
In addition to data augmentation, various sources of audio timbre can be obtained from software instruments; NSynth \cite{engel2017nsynth} is an example where the training dataset is generated from the sound of software instruments.

Pitch values tend to be continuous over time, but CREPE estimates the pitch of every frame independently without using any temporal tracking, unlike pYIN which exploits this by using an HMM to enforce temporal smoothness.
We can potentially improve the performance of CREPE even further by enforcing temporal smoothness.
In the future, we plan to do this by means of adding recurrent architecture to our model, which could be trained jointly with the convolutional front-end in the form of a convolutional-recurrent neural network (CRNN).

\linespread{0.94}\selectfont
\bibliographystyle{IEEEbib}
\bibliography{refs}

\end{document}